\documentclass[aps,longbibliography,superscriptaddress,onecolumn,showpacs,showkeys,floatfix,a4paper,eqsecnum]{revtex4}

\usepackage{textcomp}
\usepackage{eurosym}
\usepackage{amsfonts}
\usepackage{array}
\usepackage{comment}
\usepackage{amsthm}
\usepackage{changes}
\usepackage{palatino}
\usepackage[colorinlistoftodos]{todonotes}
\usepackage{mathpazo}
\usepackage{supertabular}
\usepackage{subfig}
\usepackage{amssymb}
\usepackage{eurosym}
\usepackage{amsmath}
\usepackage{graphics}
\usepackage{color}
\usepackage{graphicx}
\usepackage[colorlinks=true,
            linkcolor=blue,
            urlcolor=blue,
            citecolor=blue]{hyperref}
\def\be{\begin{equation}}

\begin{document}

\title{Gravitational Lensing Under the Effect of Weyl and Bumblebee Gravities: Applications of Gauss-Bonnet Theorem }

\author{Ali \"{O}vg\"{u}n}
\email{ali.ovgun@pucv.cl}
\homepage{https://aovgun.weebly.com} 

\affiliation{Instituto de F\'{\i}sica,
Pontificia Universidad Cat\'olica de Valpara\'{\i}so, Casilla 4950,
Valpara\'{\i}so, Chile.} 

\affiliation{Physics Department, Arts and Sciences Faculty, Eastern Mediterranean
University, Famagusta, North Cyprus via Mersin 10, Turkey.}

\author{Kimet Jusufi}
\email{kimet.jusufi@unite.edu.mk}
\affiliation{Physics Department, State University of Tetovo, Ilinden Street nn, 1200,
Tetovo, Macedonia.}
\affiliation{Institute of Physics, Faculty of Natural Sciences and Mathematics, Ss. Cyril and Methodius University, Arhimedova 3, 1000 Skopje, Macedonia.}

\author{Izzet Sakalli}
\email{izzet.sakalli@emu.edu.tr}
\affiliation{Physics Department, Arts and Sciences Faculty, Eastern Mediterranean
University, Famagusta, North Cyprus via Mersin 10, Turkey.}

\begin{abstract}
In this paper, we use the Gauss Bonnet theorem to obtain the deflection angle by the photons coupled to Weyl tensor in a Schwarzschild black hole and Schwarzschild-like black hole in bumblebee gravity in the weak limit approximation. To do so, we first calculate the corresponding optical metrics, and then we find the Gaussian curvature to use in Gauss-Bonnet theorem, which is first done by Gibbons and Werner. Hence, in the leading order terms we show the deflection angle, that is affected by the coupling between the photon and Weyl tensor, and there is a deviation from the deflecting angle as compared with Schwarzschild black hole with Schwarzschild-like black hole in bumblebee gravity. Moreover, we investigate the deflection angle by  Einstein-Rosen type wormhole in Weyl gravity and in bumblebee gravity. Interestingly, the deflection angle by Einstein-Rosen type wormhole in bumblebee gravity is found as larger than the the deflection angle by Einstein-Rosen type wormhole in Weyl gravity.

\end{abstract}



\date{\today}

\maketitle

\section{Introduction}
Over the past decade, generalized Einstein-Maxwell theories have been
receiving great attention. Those theories consider higher derivative
interactions and thus reveal more information about the features and effects
of the electromagnetic (em) fields. In general, we can split the generalized
Einstein-Maxwell theories into two classes: (i) minimally coupled gravity-em
in which there exists no coupling between the Maxwell tensor and the curvature
in the action. For example, Born-Infeld theory \cite{borninfeld} is of this class, which
eliminates the divergent self energy of the electron by modifying Maxwell's
theory and give good physical results such as the absence of shock waves and
birefringence phenomena \cite{Peres,Pellicer,Boillat}, (ii) non-minimal coupling between the
gravitational and Maxwell fields in the action \cite{Balakin,Hehl}.
Such non-minimal couplings in the Lagrangian changes the coefficients of the
second-order derivatives appeared in the Maxwell and Einstein equations.
Therefore, the propagation of gravitational and em waves in the manifold has
time delays \cite{Balakin}. In this way, the physics of the evolution of the
early Universe (quantum fluctuations of the em
fields and inflation \cite{Turner,Turner2,Mazzitelli,Mazzitelli2,Capozziello:1999uwa,Campanelli:2008qp,novello,Iorio:2013ifa,Lambiase:2005kb,ovgun1,ovgun2,ovgun3}) is expected to be explained. 

One of the generalized Einstein-Maxwell theories is the Weyl corrected
electrodynamics that involves a coupling between the Weyl tensor (WT) and the
Maxwell field \cite{Weyl1,Drummond}. Namely, the Lagrangian density of the
electrodynamics is modified with the WT. In other words, this theory \cite{Weyl1} is a special kind of em theory that it involves a coupling between
the gravitational and em fields. In fact, QED (quantum electrodynamics) of the
light effective action for one-loop vacuum polarization on a curved background
\cite{Drummond} admits such couplings of the Weyl and Maxwell tensors.
Besides, Weyl corrected electrodynamics could play a role on the supermassive
black holes located at the center of galaxies \cite{Dereli1,Solanki}. The
effects of the Weyl corrections on black hole physics are explored in
\cite{Weyl1}, which considers the holographic conductivity and diffusion in
the presence of the Weyl corrections for the AdS spacetime. It was shown that
Weyl corrections teminate the central charge seen at the leading order, tunes
the critical temperature at which holographic superconductors occur, and
modifies the order of the phase transition of the holographic superconductor
\cite{Wu:2010vr,Momeni:2011ca,Momeni:2012uc,Momeni:2012ab,Ma:2011zze,Roychowdhury:2012hp,Zhao:2012kp}. Moreover, Weyl corrections have a significant influence on
the stability of the Schwarzschild black hole \cite{Chen:2013ysa,Chen:2014zua,Chen:2015bva,Chen:2015cpa,Lu:2016gsf,Huang:2016qnl,Chen:2016hil,Li:2017zsb,Cao:2018lrd}.

In addition to all these, the bumblebee gravity model to be used in this study is dynamically Lorentz
symmetry breaking (LSB) in terms of charge conjugation, parity transformation, and time
reversal \cite{Bluhm:2004ep,Bertolami:2005bh,Bluhm:2008yt,Seifert:2009gi,Escobar:2017fdi}. By defining the bumblebee vector, the model also acquires rotation and boost properties. In 1989, Kostelecky and Samuel showed that mechanisms arising in the context of string theory can lead to the LSB \cite{Kostelecky:1989jw,namb}. The motivation of the bumblebee mechanism came from the string theory and give a spontaneous LSB by tensor-valued fields acquiring vacuum expectation values \cite{Kostelecky:2010ze}. Recently, the exact solution of the Schwarzschild-like bumblebee black hole has been derived \cite{bumblebee}. In fact, this kind of gravity model, which admits LSB property has been extensively studied in the literature \cite{LSB,Maluf:2015hda,Capelo:2015ipa,Nascimento:2014vva,Maluf:2014dpa,Paramos:2014mda}. One of the main advantages of taking into account of the LSB is to reveal the effects of string and loop quantum gravity theories at low energy levels.

A wormhole is a theoretical passage through spacetime that could help people and things travel huge distances through space in short amounts of time. Albert Einstein and Nathan Rosen \cite{Einstein:1935tc} proposed this theory in 1935; wormholes are also known as Einstein-Rosen bridges. According to Einstein's theory of general relativity, they mathematically should exist. A wormhole has two mouths connected by a throat that connects two different points in spacetime. They may not only connect two space points, but they might also be able to connect two universes. So one can ask; could we travel in a wormhole like plenty of science-fiction movies suggest? According to Einstein and Rosen's theory, which states that a wormhole collapses quickly, the answer is no. On the other hand, new theories suggest that wormholes may stay open longer with the existence of "exotic matter". But it seems that we are currently far from having the technology required to find and use the exotic matter. However, those kind of mysterious objects might be detected in the future by the help of gravitational lensing. In the
recent years, the most commonly used method to investigate the weak
gravitational lensing is the Gauss-Bonnet theorem \cite{gauss,bonnet,GBT} or the so-called the
Gibbons-Werner method (GWM) \cite{gibbons1,gibbons2,Gibbons:2015qja}. This method has brought dramatic ease and enhancement to the calculation of gravitational lensing. Gibbons and
Werner proved that by applying the Gauss-Bonnet theorem to the corresponding
optical metric of the spacetime considered, one can straightforwardly compute
the deflection angle. The latter remark highlights that the bending of light
ray has a global effect which is contrary to the popular opinion. Because,
the bending of light is usually computed for a compact region having a radius
at the order of the impact parameter. GWM indeed focuses on a non-singular
domain, which is outside of the light ray. 

In the GWM, one use the optical geometry and then calculate the Gaussian optical curvature $K$ to find the asymptotic bending angle which can be calculated as follows \cite{gibbons1,werner}:
\begin{equation}
\hat{\alpha}=-\int \int_{D_\infty} K  \mathrm{d}S,
\end{equation} which yields exact result for the bending angle.

In addition to this, the original GWM
computes the deflection angle for the asymptotically flat spacetimes. For
non-asymptotically flat (NAF) spacetimes, the GWM becomes valid when taking cognizance of the finite distance corrections \cite{gibbons1,asahi1,Ono:2017pie}. Soon after, Werner \cite{werner} extended the
GWM to cover rotating black holes (Kerr) in which Finsler-Randers metric was considered. Then, the GWM was studied for the gravitational lensing problems in the rotating/non-rotating geometries of wormholes \cite{Jusufiwh,Jusufi:2017mav}. Today, the GWM has been employed in the numerous studies (see for
example
\cite{Ovgun:2018xys,asahi1,Ono:2017pie,  Jusufiwh,Jusufi:2017mav,kimet2,kimet22,wh10,kimet1,kaa,kimet3,aovgun,aovgun1,asada,Crisnejo:2018uyn,Jusufi:2018jof,Jusufi:2018kmk,Jusufi:2017drg,Jusufi:2017uhh,Jusufi:2017xnr,Jusufi:2017vew,Ishihara:2016vdc,Jusufi:2015laa,Gibbons:2015qja}).

In this paper, our main motivation is to explore
the effects of the extended gravitational theories on the gravitational
lensing. Nowadays, gravitational lensing for photons coupled to the WT has gained much attention because of the supermassive black hole at the Galactic center, Sgr A*, which is expected to be photographed by the Event Horizon Telescope \cite{eventhorizon}. For this purpose, we consider the Schwarzschild black hole with Weyl
\cite{Weyl1} and bumblebee \cite{bumblebee} corrections. As aforementioned above, the bumblebee gravity is an effective field
theory that describes a vector field with a vacuum expectation value that
spontaneously breaks Lorentz symmetry \cite{Kostelecky:1989jw,Kostelecky:1989jp,Kostelecky:1988zi}. We shall\ apply the GWM to those black holes and analyze the effects of the Weyl and bumblebee corrections
on their gravitational lensing. 

The paper is organized as follows: in the following section we introduce the
Weyl corrected Schwarzschild black hole and derive the deflection angle in the
context of the GWM. In Sec.III, the modified Schwarzschild black hole in the
bumblebee gravity is summarized and study the change of the deflection angle
due to the bumblebee corrections. Moreover, Einstein-Rosen type wormhole
solutions and their gravitational lensings are thoroughly studied in Secs. IV
and V, respectively. Finally, Sec. VI concludes the paper.

\section{Weyl correction of a Schwarzschild black hole and weak gravitational lensing}
In this section, to study the effect of the Weyl correction on the deflection angle we use following the action of the Einstein-Maxwell theory, which is coupled to the WT in the 4-dimensional static and spherical
symmetric spacetime \cite{Weyl1}:
\begin{equation}
S=\int d^{4}x\sqrt{-g}\bigg[\frac{R}{16\pi G}-\frac{1}{4}\bigg(F_{\mu\nu}F^{\mu\nu}-4\alpha C^{\mu\nu\rho\sigma}F_{\mu\nu}F_{\rho\sigma}\bigg)\bigg],\label{acts}
\end{equation}
where  $C_{\mu\nu\rho\sigma}$ stands for the WT as 

\begin{equation}
C_{\mu\nu\rho\sigma}=R_{\mu\nu\rho\sigma}-\frac{1}{2}(g_{\mu[\rho}R_{\sigma]\nu}-g_{\nu[\rho}R_{\sigma]\mu})+\frac{Rg_{\mu[\rho}g_{\sigma]\nu}}{6},
\end{equation}
where the square brackets around multiple indices denote the antisymmetrized part of the tensor. Moreover, the electromagnetic tensor $F_{\mu\nu}$
is equal to $F_{\mu\nu}=A_{\nu;\mu}-A_{\mu;\nu}$. Note that
 $\alpha$ is the coupling constant with a dimension of length-squared.
 
The equation of motion for photon coupled to the WT is found as
\begin{eqnarray}
k_{\mu}k^{\mu}a^{\nu}+8\alpha C^{\mu\nu\rho\sigma}k_{\sigma}k_{\mu}a_{\rho}=0.\label{WE2}
\end{eqnarray}
Clearly, propagation
of the coupled photon is affected by coupling term  $\alpha$  with the WT. Hence, the coupled photons move non-geodesically in the curved spacetime. Normally, it is known that photons should follow null geodesics  $\gamma_{\mu\nu}$, i.e., $\gamma^{\mu\nu}k_{\mu}k_{\nu}=0$
\cite{Breton}. Using the Einstein field equations with above equations, one can define the effective metric for the coupled photon as follows \cite{Chen:2015cpa}:
\begin{eqnarray}
ds^{2}=-A(r)dt^{2}+A(r)^{-1}dr^{2}+C(r)W(r)^{-1}d\Omega^2.\label{l01} \label{wyl}
\end{eqnarray}
Note that $d\Omega^{2}=d\theta^{2}+\sin^{2}\theta d\phi^{2}$ and $A(r)=1-\frac{2M}{r}$ and $C(r)=r^{2}$. Furthermore, for two different polarizations of the photon respectively along $ l_\mu$ $(PPL)$ and  $m_\mu$ $(PPM)$  the quantity 
$W(r)$ are 

\begin{eqnarray}
W(r)_{PPL}=\frac{r^{3}-8\alpha M}{r^{3}+16\alpha M},\label{v12}
\end{eqnarray}
and
\begin{eqnarray}
W(r)_{PPM}=\frac{r^{3}+16\alpha M}{r^{3}-8\alpha M}.\label{v12}
\end{eqnarray}

Assuming both the observer and the source are located in the equatorial plane as well as the trajectory of the null photon is restricted on the same plane with ($\theta=\frac{\pi}{2}$),
the metric is reduced to the following form:
\begin{eqnarray}
ds^{2}=-A(r)dt^{2}+A(r)^{-1}dr^{2}+C(r)W(r)^{-1}d\phi^{2}.\label{l1}
\end{eqnarray}
For the photon moving in the equatorial plane ($\theta=\frac{\pi}{2}$),
we have $k_{2}=0$ and the null geodesics,
$\mathrm{d}s^{2}=0$, result in
\begin{eqnarray}
\mathrm{d}t^{2}=\frac{\mathrm{d}r^{2}}{A(r)^{2}}+\frac{C(r)\mathrm{d}\varphi^{2}}{W(r)A(r)}.\label{metric2}
\end{eqnarray}

In terms of the new coordinate $r^{\star}$, the optical metric tensor $\tilde{g}_{ab}$ satisfies 
\begin{equation}
\mathrm{d}t^{2}=\tilde{g}_{ab}\,\mathrm{d}x^{a}\mathrm{d}x^{b}=\mathrm{d}{r^{\star}}^{2}+f^{2}(r^{\star})\mathrm{d}\varphi^{2},\label{optical}
\end{equation}
where the function $f(r^{\star})$ is given by 
\begin{equation}
f(r^{\star})=\frac{\sqrt{C(r)}}{\sqrt{W(r)A(r)}}.\label{f}
\end{equation}
Note that determinant is equal to $\det\tilde{g}_{ab}=f^{2}(r^{\star})$. Using
the optical metric \eqref{optical}, the only non-vanishing Christoffel symbols are $\Gamma_{\varphi\varphi}^{r}=-f(r^{\star})f'(r^{\star})$
and $\Gamma_{r\varphi}^{\varphi}=f'(r^{\star})/f(r^{\star})$. Thus, one can calculate the Gaussian optical curvature
$K$ \cite{gibbons1} as follows
\begin{equation}
K=-\frac{R_{r\varphi r\varphi}}{\det\tilde{g}_{r\varphi}}=-\frac{1}{f(r^{\star})}\frac{\mathrm{d}^{2}f(r^{\star})}{\mathrm{d}{r^{\star}}^{2}}.
\end{equation}

Optical curvature $K$ can also be rewritten in terms of $r$ \cite{bao}. Thus, we have
\begin{eqnarray}
K & = &-\frac{1}{f(r^{\star})}\left[\frac{\mathrm{d}r}{\mathrm{d}r^{\star}}\frac{\mathrm{d}}{\mathrm{d}r}\left(\frac{\mathrm{d}r}{\mathrm{d}r^{\star}}\right)\frac{\mathrm{d}f}{\mathrm{d}r}+\left(\frac{\mathrm{d}r}{\mathrm{d}r^{\star}}\right)^{2}\frac{\mathrm{d}^{2}f}{\mathrm{d}r^{2}}\right],\label{Gcurvature}
\end{eqnarray}
which yields
\begin{widetext}
\begin{eqnarray}
K=-{\frac{\left(C''\right)\left(A\right)^{2}}{2C}}+\left(\frac{A''}{2}\right)A+{\frac{\left(W''\right)\left(A\right)^{2}}{2W}}+{\frac{\left({}C'\right)^{2}\left(A\right)^{2}}{4\left(C\right)^{2}}}\\+{\frac{\left({}C'\right)\left({}W'\right)\left(A \right)^{2}}{2C W}}-\left(4A'\right)^{2}-{\frac{2\left(W' \right)^{2}\left(A \right)^{2}}{4\left(W\right)^{2}}}. \notag \label{curvature}
\end{eqnarray}
We first consider the $PPL$ case. After substituting Eq. \eqref{f} into Eq. \eqref{curvature}, the optical curvature is obtained as follows:
\begin{equation}
K_{PPL}=\frac{M}{\left(-{r}^{3}+8\,\alpha\,M\right)^{2}{r}^{4}\left({r}^{3}+16\,\alpha\,M\right)^{2}}
\notag \end{equation}
\begin{equation}
\times(3\,M{r}^{12}-2\,{r}^{13}+336\,{M}^{2}\alpha\,{r}^{9}-320\,M\alpha\,{r}^{10}+72\,\alpha\,{r}^{11}-6912\,{M}^{3}{\alpha}^{2}{r}^{6}+6720\,{M}^{2}{\alpha}^{2}{r}^{7} \end{equation}
\begin{equation}
-1584\,M{\alpha}^{2}{r}^{8}+67584\,{M}^{4}{\alpha}^{3}{r}^{3}-69632\,{M}^{3}{\alpha}^{3}{r}^{4}+18432\,{M}^{2}{\alpha}^{3}{r}^{5}+49152\,{M}^{5}{\alpha}^{4}-32768\,{M}^{4}{\alpha}^{4}r),\notag
\end{equation}
\end{widetext}
which asymptotically behaves as
\begin{equation}
K_{PPL}\simeq -\frac{2M}{r^3}+\frac{3M^2}{r^4}-\frac{72 M \alpha}{r^5}.
\end{equation}
So, one can deduce from the above result how the coupling term $\alpha$ of the WT affects the optical curvature $K_{PPL}$.
\subsection{Calculation of Deflection angle}

Now, we calculate the deflection angle using the Gauss-Bonnet theorem. This theorem provides relation between the intrinsic geometry of the spacetime and its topology of the region $D_{R}$ in $M$, with boundary $\partial D_{R}=\gamma_{\tilde{g}}\cup C_{R}$
\cite{gibbons1}:
\begin{equation}
\int\limits _{D_{R}}K\,\mathrm{d}S+\oint\limits _{\partial D_{R}}\kappa\,\mathrm{d}t+\sum_{i}\epsilon_{i}=2\pi\chi(D_{R}),
\end{equation}
where $\kappa$ denotes the geodesic curvature,
given by $\kappa=\tilde{g}\,(\nabla_{\dot{\gamma}}\dot{\gamma},\ddot{\gamma})$,
such that $\tilde{g}(\dot{\gamma},\dot{\gamma})=1$, with the unit
acceleration vector $\ddot{\gamma}$, and $\epsilon_{i}$ corresponds to the
exterior angle at the $i^{th}$ vertex. As $R\to\infty$, both jump
angles become $\pi/2$, so that we get $\theta_{O}+\theta_{S}\to\pi$.
Since $D_{R}$ is non-singular, than the Euler characteristic is $\chi(D_{R})=1$,
hence we have
\begin{equation}
\iint\limits _{D_{R}}K\,\mathrm{d}S+\oint\limits _{\partial D_{R}}\kappa\,\mathrm{d}t +\theta_{i}=2\pi\chi(D_{R}).\label{gaussbonnet}
\end{equation}

$\gamma_{\tilde{g}}$ is a geodesic and $\theta_{i}=\pi$ denotes the total jump angle; thus we have $\kappa(\gamma_{\tilde{g}})=0$. Since the Euler characteristic number $\chi$ is 1, the remaining part yields $\kappa(C_{R})=|\nabla_{\dot{C}_{R}}\dot{C}_{R}|$
as $R\to\infty$. The radial component of the geodesic curvature is given by
\begin{equation}
\left(\nabla_{\dot{C}_{R}}\dot{C}_{R}\right)^{r}=\dot{C}_{R}^{\varphi}\,\partial_{\varphi}\dot{C}_{R}^{r}+\Gamma_{\varphi\varphi}^{r}\left(\dot{C}_{R}^{\varphi}\right)^{2}. \label{izo1}
\end{equation}
At very large $R$, $C_{R}:=r(\varphi)=R=const.$ Thus, the first term of Eq. (\ref{izo1}) becomes zero and $(\dot{C}_{R}^{\varphi})^2=1/f^{2}(r^{\star})$. Recalling $\Gamma_{\varphi\varphi}^{r}=-f(r^{\star})f'(r^{\star})$, we have
\begin{equation}
\left(\nabla_{\dot{C}_{R}^{r}}\dot{C}_{R}^{r}\right)^{r}\to-\frac{1}{R},
\end{equation}
and it follows that the geodesic
curvature is independent of topological defects, $\kappa(C_{R})\to R^{-1}$. However from the optical metric \eqref{optical}, it is not difficult
to see that $\mathrm{d}t=\,R\,\mathrm{d}\,\varphi$. Whence, one gets 
\begin{equation}
\kappa(C_{R})\mathrm{d}t=\frac{1}{R}\,R\,\mathrm{d}\,\varphi.
\end{equation}
Taking cognizance of the above results, we obtain
\begin{eqnarray}
\iint\limits _{D_{R}}K\,\mathrm{d}S & + & \oint\limits _{C_{R}}\kappa\,\mathrm{d}t\overset{{R\to\infty}}{=}\iint\limits _{S_{\infty}}K\,\mathrm{d}S+\int\limits _{0}^{\pi+\hat{\alpha}}\mathrm{d}\varphi.\label{def1}
\end{eqnarray}

In the weak deflection limit,  we may assume that the light ray is given
by $r(t)=b/\sin\varphi$ at zeroth order. Using \eqref{curvature}
and \eqref{def1}, the deflection angle thus becomes
\begin{equation}
\hat{\alpha}=-\int\limits _{0}^{\pi}\int\limits _{r_\gamma}^{\infty}K\,\sqrt{\det\bar{g}}\,\mathrm{d}r^{\star}\,\mathrm{d}\varphi,\label{angle}
\end{equation}
where
\begin{eqnarray}
\sqrt{\det\tilde{g}}\mathrm{d}r^{\star}=r\mathrm{d}r\left(1+\frac{3M}{r}+....\right),
\end{eqnarray}
and $r_\gamma$ is given by \cite{Ono:2017pie}
\begin{eqnarray}
\frac{1}{r_\gamma}&=&\frac{\sin\varphi}{b}+\frac{M(3+\cos(2\varphi))}{2b^2}\\\notag
&+&\frac{M^2(37\sin\varphi+30(\pi-2\varphi)\cos\varphi-3\sin(3\varphi))}{16 b}.
\end{eqnarray}

Substituting the leading order terms of the Gaussian curvature \eqref{curvature}
into the last equation, we find the deflection angle up to second order terms as follows:

\begin{eqnarray}
\hat{\alpha}_{PPL} & \approx & \int\limits _{0}^{\pi}\int\limits _{r_\gamma}^{\infty}\left(-\frac{2M}{r^3}+\frac{3M^2}{r^4}-\frac{72 M \alpha}{r^5}\right)\sqrt{\det\tilde{g}}\mathrm{d}r^{\star}\mathrm{d}\varphi\nonumber \\
 & \approx & \frac{4M}{b}+\frac{15 \pi M^2}{4b^2}+\frac{32 M \alpha}{b^3}+\frac{261 \pi M \alpha }{4 b^4}. \label{myi1}
\end{eqnarray}

As a second case, we consider the $PPM$ case, which admits the following Gaussian curvature 
\begin{equation}
K_{PPM}\simeq -\frac{2M}{r^3}+\frac{3M^2}{r^4}+\frac{72 M \alpha }{r^5}.
\end{equation}

After some algebra, we also obtain the following deflection angle:
\begin{eqnarray}
\hat{\alpha}_{PPM} & \approx & \int\limits _{0}^{\pi}\int\limits _{\frac{b}{\sin\varphi}}^{\infty}(-\frac{2M}{r^3}+\frac{3M^2}{r^4}+\frac{72 M \alpha}{r^5})\sqrt{\det\tilde{g}}\mathrm{d}r\mathrm{d}\varphi\nonumber \\
 & \approx & \frac{4M}{b}+\frac{15 \pi M^2}{4b^2}-\frac{32 M \alpha}{b^3}-\frac{261 \pi M \alpha }{4 b^4}.\label{myi2}
\end{eqnarray}

Thus, one can immediately observe how the coupling term $\alpha$ of the WT influences the deflection angles (\ref{myi1}) and (\ref{myi2}).
 
 \section{Deflection angle of Schwarzschild-like solution in a bumblebee gravity}

The metric of Schwarzschild-like solution in a bumblebee gravity spherically is given by
\cite{bumblebee}
\begin{equation}
ds^{2} =-\left( 1-\frac{2M}{r}\right) dt^{2}+\left( 1+\ell \right) \left(
1-\frac{2M}{r}\right) ^{-1}dr^{2}+r^{2}d\theta ^{2}+r^{2}\sin ^{2}\theta d\phi ^{2}\text{,}  \label{line}
\end{equation}%
where we have conveniently identified $\rho _{0}\equiv
2M $ ($M=G_{N}m$ is the usual geometrical mass). $\ell=\xi b^{2}$ ($\xi:const.$) \cite{bumblebee} such that in the limit $\ell
\rightarrow 0~(b^{2}\rightarrow 0)$, we recover the usual Schwarzschild
metric. The metric (\ref{line}) represents a purely radial LSB solution
outside a spherical body characterizing a modified black
hole solution. 

The null geodesics of the metric (\ref{line}) leads to the following optical line-element for the Schwarzschild-like solution in a bumblebee gravity:
\begin{eqnarray}
\mathrm{d}t^{2}=\frac{\left( 1+\ell \right)\mathrm{d}r^{2}}{\left(
1-\frac{2M}{r}\right)^{2}}+\frac{r^2\mathrm{d}\varphi^{2}}{\left(
1-\frac{2M}{r}\right)}.\label{metric2}
\end{eqnarray}

The metric tensor of the above optical metric $\tilde{g}_{ab}$ can be written
in terms of the new coordinate $r^{\star}$ as follows
\begin{equation}
\mathrm{d}t^{2}=\tilde{g}_{ab}\,\mathrm{d}x^{a}\mathrm{d}x^{b}=\mathrm{d}{r^{\star}}^{2}+f^{2}(r^{\star})\mathrm{d}\varphi^{2},\label{optical}
\end{equation}
where the function $f(r^{\star})$ reads
\begin{equation}
f(r^{\star})=\frac{r}{\sqrt{\left(
1-\frac{2M}{r}\right)}}.\label{f}
\end{equation}
 Using Eq. \eqref{optical}, we compute the Gaussian optical curvature for the Schwarzschild-like solution in the bumblebee gravity
$K_{BGSCH}$ \cite{gibbons1}:
\begin{equation}
K_{BGSCH}={\frac {M \left( 3\,M-2\,r \right) }{ \left( 1+l \right) {r}^{4}}},
\end{equation}
which approximates to the following form at asymptotic region
\begin{equation}
K_{BGSCH}\simeq -\,{\frac {2M}{ \left( 1+l \right) {r}^{3}}}.
\end{equation}

The geodesic curvature is then found to be 

\begin{equation}
\kappa \to \frac{1}{R \sqrt{1+l}},
\end{equation}
which results in
\begin{equation}
\kappa dt \to \frac{1}{ \sqrt{1+l}}d\varphi.
\end{equation}

It follows that the deflection angle is given by
\begin{equation}
\hat{\alpha}=\pi\left( \sqrt{1+l}-1\right)- \sqrt{1+l}\int\limits _{0}^{\pi}\int\limits _{r_\gamma}^{\infty}K\,\sqrt{\det\bar{g}}\,\mathrm{d}r^{\star}\,\mathrm{d}\varphi.\label{angle}
\end{equation}
With the aid of the below expansion
\begin{eqnarray}
 \sqrt{1+l}=1+\frac{ l}{2}+...,
\end{eqnarray}

we find the deflection angle of the gravitational lensing of the Schwarzschild-like solution in the bumblebee gravity as follows
\begin{eqnarray}
\hat{\alpha}_{BSCH} & \approx & \frac{\pi l}{2} +\frac{4M}{b}-\frac{2 M l}{b}.
\end{eqnarray}

\section{Lensing by Einstein-Rosen type wormhole in Weyl gravity}

We now introduce a new coordinate transformation $u^{2}=r-2M$ to the spacetime metric (\ref{wyl}) to remove the singularity and transform it to a Einstein-Rosen like wormhole in Weyl gravity (WERW), as follows: 
\begin{eqnarray}
ds^{2}=-\frac{u^2}{u^2+2M}dt^{2}+4(u^2+2M)du^{2}+\frac{(u^2+2M)^2}{B(u)}d\Omega^2, \notag \\ B(u)=\frac{(u^2+2M)^3-8\alpha M}{(u^2+2M)^3+16\alpha M} \notag.\end{eqnarray}

Note that the radius of the throat is located at $u_{throat}=0$ and it is non-singular in the interval of $u\in (-\infty ,\infty )$. Then we write the optical metric of WERW as follows:
\begin{equation}
\mathrm{d}t^{2}=\frac{4(u^2+2M)^2 }{u^{2}}\mathrm{d}u^{2}+\frac{(u^2+2M)^3}{u^{2} B(u) }\mathrm{d}\varphi ^{2}.  \label{63}
\end{equation}

The Gaussian optical curvature is calculated as:
\begin{equation}
\mathcal{K}\simeq \frac{1}{4{u}^{4}}-4\,{\frac { \left( {u}^{4}+27\,\alpha \right) M}{{u}^{10
}}}+{\frac { \left( 25\,{u}^{8}+1512\,\alpha\,{u}^{4}+432\,{\alpha}^{2
} \right) {M}^{2}}{{u}^{16}}}
 .  
\end{equation}

Hence, the deflection angle of WERW is found to be
\begin{equation}
\hat{\alpha}_{WERW}\simeq \,{\frac {\pi}{16{b}^{2}}}+\,{\frac {3\pi\,M}{8{b}^{4}}}
.
\end{equation}

\section{Lensing by Einstein-Rosen type wormhole in Bumblebee gravity}
Similar to the previous section,  we apply a new coordinate transformation $u^{2}=r-2M$ to the metric (\ref{line}) in order to remove the singularity and convert it to a Einstein-Rosen wormhole of the bumblebee gravity (BERW). Thus, we have 
\begin{eqnarray}
ds^{2}=-\frac{u^2}{u^2+2M}dt^{2}+\left( 1+\ell \right)4(u^2+2M)du^{2}+ (u^2+2M)^2d\Omega^2.\end{eqnarray}

Note that the radius of the throat is located at $u_{throat}=0$ and non-singular in the interval $u\in (-\infty ,\infty )$.
Then, one can express the optical metric of the BERW as follows:
\begin{equation}
\mathrm{d}t^{2}=\left( 1+\ell \right)\frac{4(u^2+2M)^2 }{u^{2}}\mathrm{d}u^{2}+\frac{(u^2+2M)^3}{u^{2} }\mathrm{d}\varphi ^{2}.  \label{63}
\end{equation}

The Gaussian optical curvature can be computed as

\begin{equation}
\mathcal{K}={\frac {{u}^{4}-8\,M{u}^{2}-4\,{M}^{2}}{4 \left( {u}^{2}+2\,M
 \right) ^{4} \left( 1+l \right) }}
\end{equation}
which asymptotically behaves as 
\begin{equation}
\mathcal{K}\simeq \,{\frac {1}{4{u}^{4} \left( 1+l \right) }}-4\,{\frac {M}{{u}^{6}
 \left( 1+l \right) }}+25\,{\frac {{M}^{2}}{{u}^{8} \left( 1+l
 \right) }}
 .  
\end{equation}

The geodesic curvature is obtained as
\begin{equation}
\kappa \to \frac{1}{u^2 \sqrt{1+l}}.
\end{equation}
Hence, we have
\begin{equation}
\kappa dt \to  u^2\,d\varphi
\end{equation}
which yields
\begin{equation}
\kappa dt \to \frac{1}{ \sqrt{1+l}}d\varphi.
\end{equation}
Finally the deflection angle becomes
\begin{equation}
\hat{\alpha}_{BERW}=\pi\left( \sqrt{1+l}-1\right)- \sqrt{1+l}\int\limits _{0}^{\pi}\int\limits _{r_\gamma}^{\infty}K\,\sqrt{\det\bar{g}}\,\mathrm{d}r^{\star}\,\mathrm{d}\varphi,
\end{equation}
in which $\sqrt{\det\bar{g}}\mathrm{d}r^{\star}\,\mathrm{d}\varphi \simeq 2 \sqrt{1+l}\,u^3 du d\varphi$. Hence, the deflection angle of the BERW is found to be as follows:
\begin{equation}
\hat{\alpha}_{BERW} \simeq \frac{\pi l}{2}+ \,{\frac {\pi}{16{b}^{2}}}+\,{\frac {3\pi\,M}{8
{b}^{4}}}.
\end{equation}

\section{Conclusion}
In this paper, we have studied the effect of the Weyl corrections by using the photons, which are coupled to the WT in a Schwarzschild black hole in two different cases: PPL and PPM. For this purpose, we have used the GWM to calculate the weak gravitational lensing. The following deflection angles are obtained:

\begin{eqnarray}
\hat{\alpha}_{PPL} & \approx &  \frac{4M}{b}+\frac{15 \pi M^2}{4b^2}+\frac{32 M \alpha}{b^3}+\frac{261 \pi M \alpha }{4 b^4},
\end{eqnarray}
and

\begin{eqnarray}
\hat{\alpha}_{PPM} & \approx & \frac{4M}{b}+\frac{15 \pi M^2}{4b^2}-\frac{32 M \alpha}{b^3}-\frac{261 \pi M \alpha }{4 b^4}.\label{angle2}
\end{eqnarray}

\begin{figure}[!ht]
\centering
\includegraphics[width=0.4\textwidth]{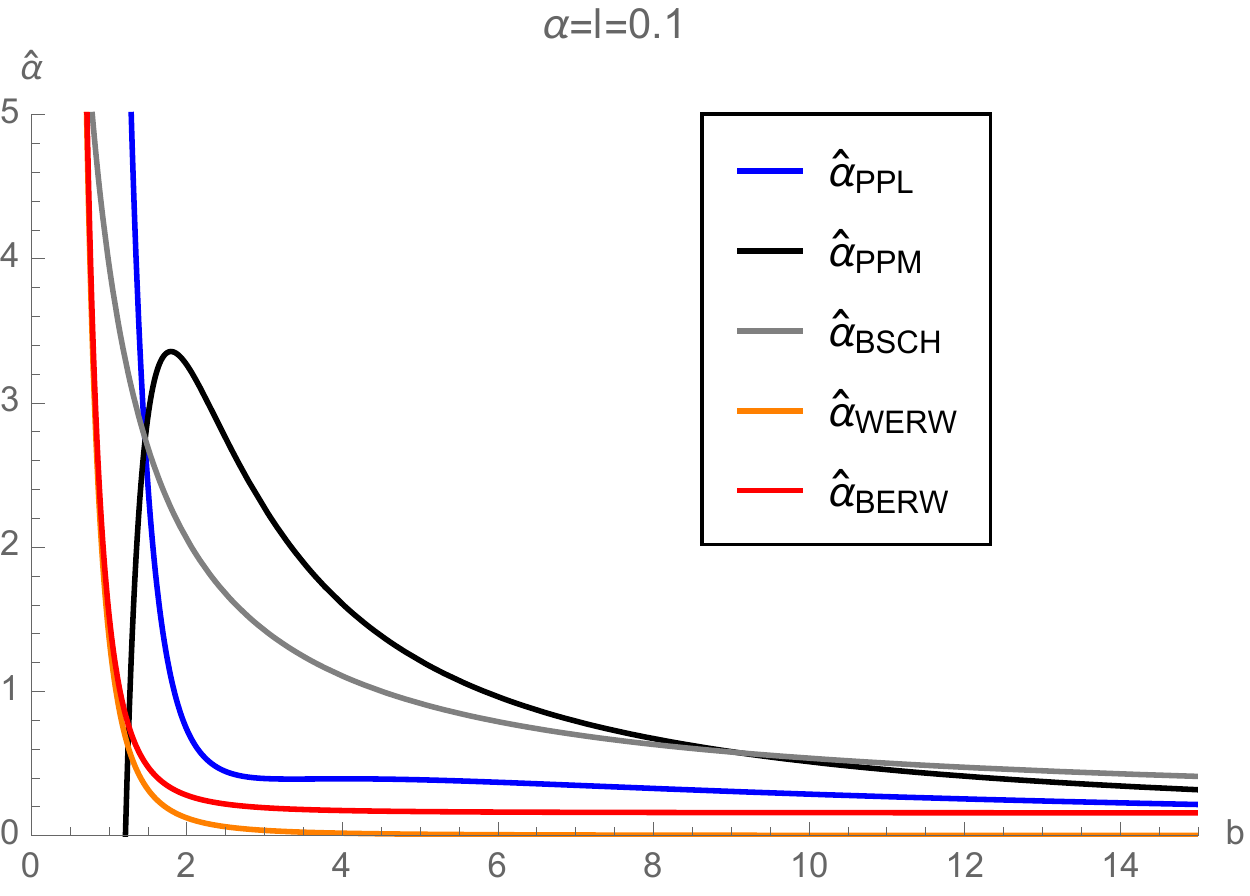}
\includegraphics[width=0.4\textwidth]{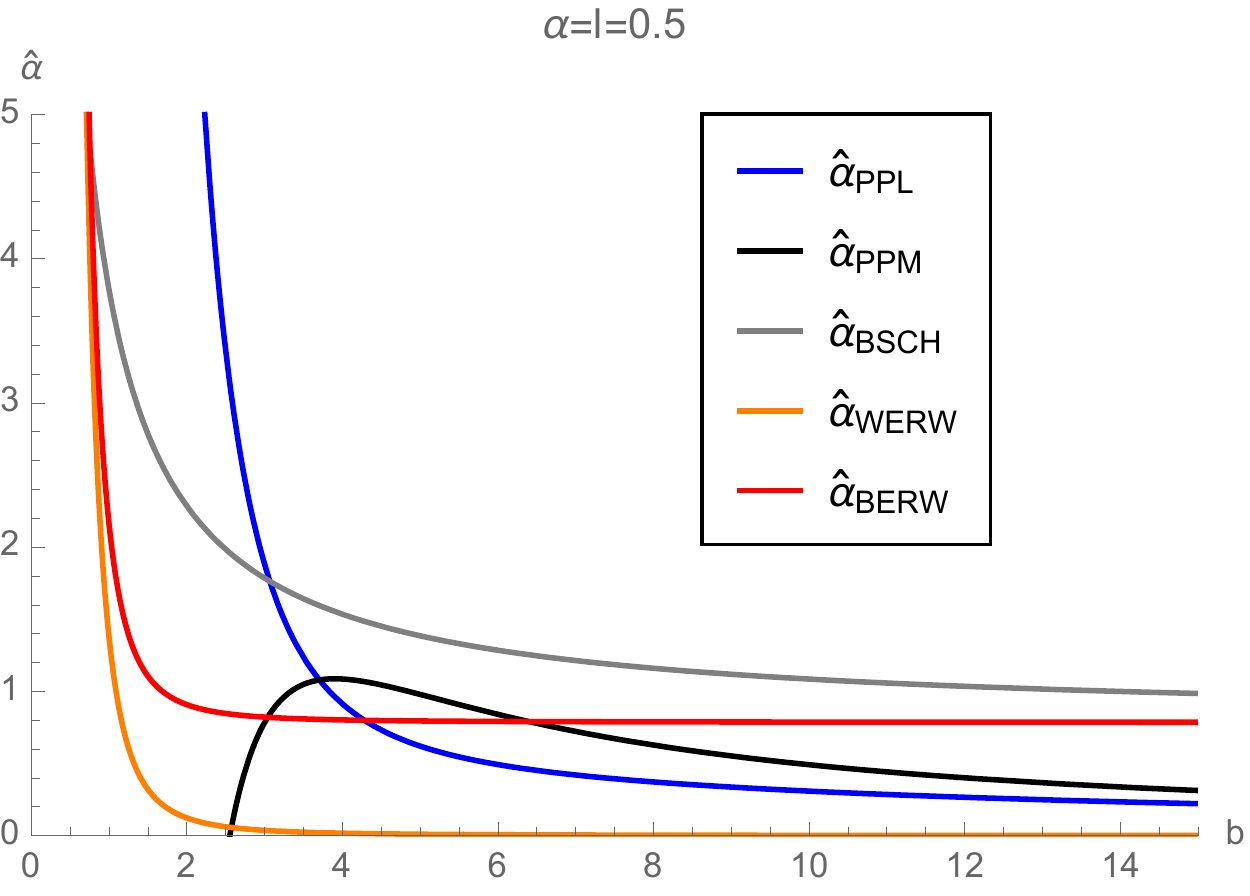}
\includegraphics[width=0.4\textwidth]{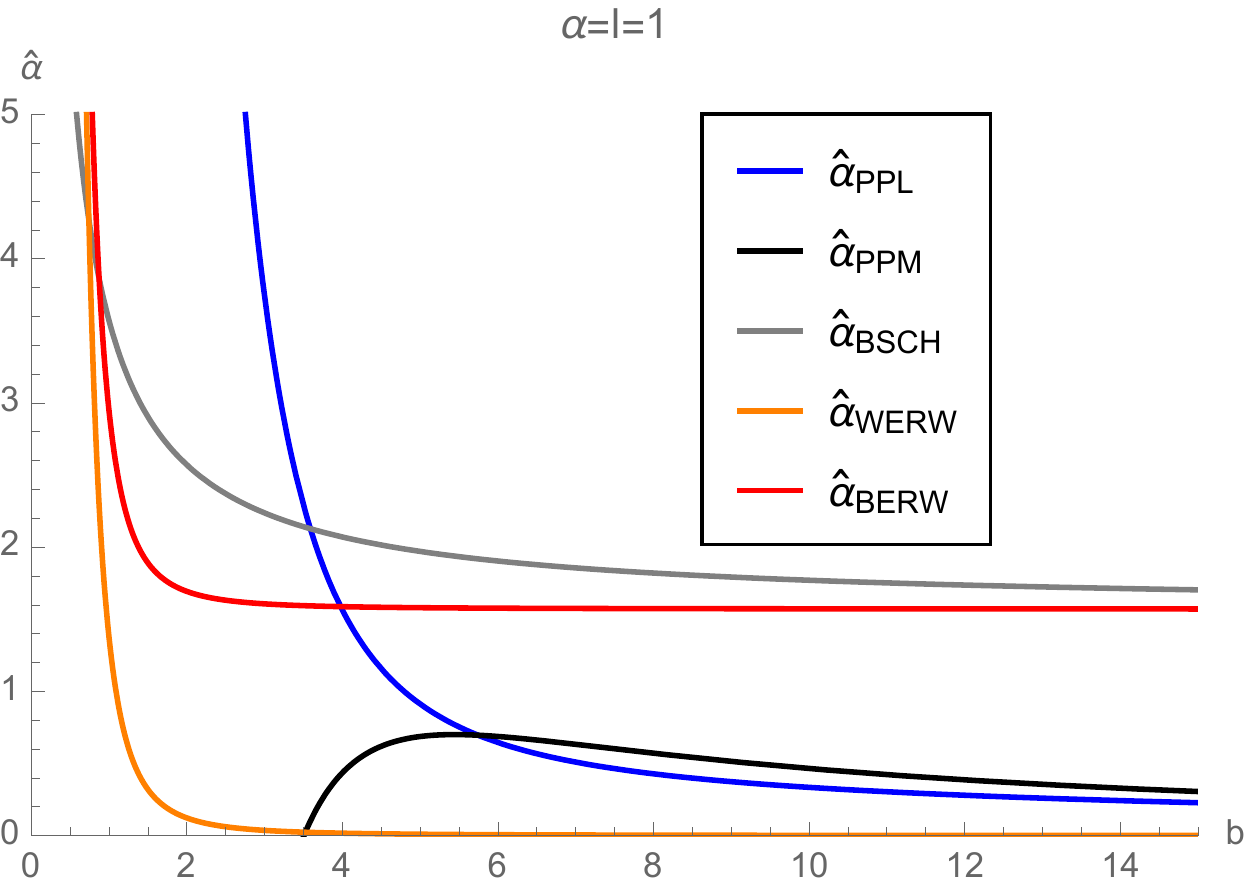}
\includegraphics[width=0.4\textwidth]{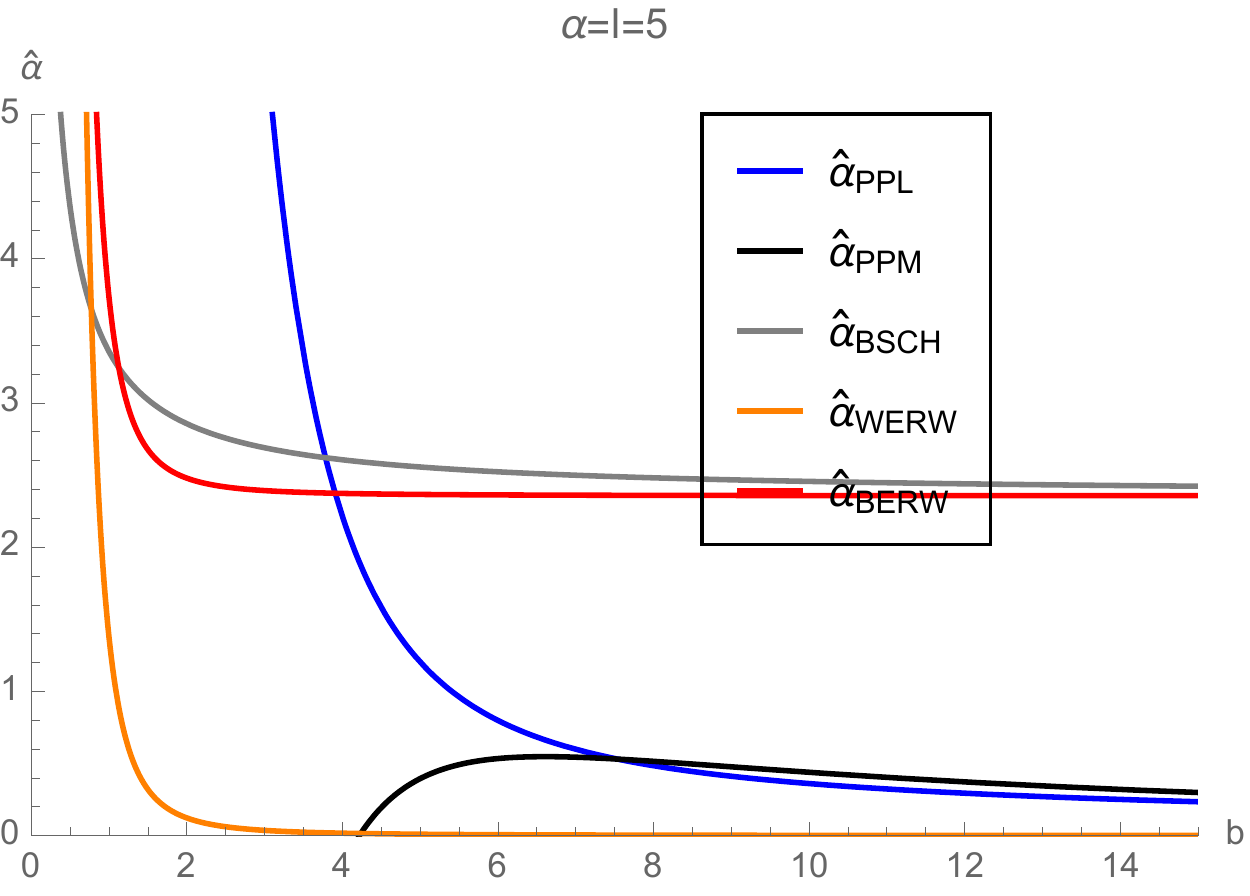}
\caption{The deflection angle of $\hat{\alpha}_{PPL}$, $\hat{\alpha}_{PPM}$, $\hat{\alpha}_{BSCH}$, $\hat{\alpha}_{WERW}$ and $\hat{\alpha}_{BERW}$ versus impact parameter $b$ with variable constants $\alpha$ and $l$. } \label{Fig1}
\end{figure}

Then, we have applied the GBT to Schwarzschild-like black hole in bumblebee gravity and obtained its total deflection angle:
\begin{eqnarray}
\alpha_{BSCH} & \approx & \frac{\pi l}{2} +\frac{4M}{b}-\frac{2 M l}{b}.
\end{eqnarray}

Afterward, we have constructed Einstein-Rosen bridges for the Schwarzschild-like black holes originated from the WT and bumblebee gravity. Their deflection angles are found to be
\begin{equation}
\hat{\alpha}_{WERW}\simeq \,{\frac {\pi}{16{b}^{2}}}+\,{\frac {3\pi\,M}{8{b}^{4}}},
\end{equation}
and 
\begin{equation}
\hat{\alpha}_{BERW} \simeq \frac{\pi l}{2}+ \,{\frac {\pi}{16{b}^{2}}}+\,{\frac {3\pi\,M}{8
{b}^{4}}}.
\end{equation}

In conclusion, we have managed to show how the effect of Weyl and bumblebee parameters influence the deviation of deflection: see Fig. \ref{Fig1}.

Using the units based on Sun as $m=M_{\odot }$ and $D\approx R_{\odot }$ with the values in Ref. \cite{Beringer}, it is easy to show that an angle predicted by theory of the general relativity is $\hat{\alpha} _{\text{GR}}=4G_{N}M_{\odot
}/c^{2}R_{\odot }\approx {{}1.7516687}^{\prime \prime }$.  Consequently, the expected value of the deflection angle due to the effect of the photons coupled to either WT or LSB would be higher than $1.7516687^{\prime \prime}$.

\acknowledgments
 We wish to thank the Editor and anonymous Referee for their valuable comments and
suggestions. This work was supported by the Chilean FONDECYT
Grant No. 3170035 (A. \"{O}.).


\begin{thebibliography}{1d1}
\bibliographystyle{prd}




\bibitem{borninfeld}
 M. Born and L. Infeld, Proc Roy Soc (Lond) A144 425 (1934): a useful review of the theory may be found in M Born, Ann Inst Poincare 7 155 (1939).
 
 \bibitem{Peres} 
  A.~Peres,
  Phys.\ Rev.\  {\bf 122}, 273 (1961).
  
  \bibitem{Pellicer} 
  R.~Pellicer and R.~J.~Torrence,
  J.\ Math.\ Phys.\  {\bf 10}, 1718 (1969)..
 

\bibitem{Boillat} 
  G.~Boillat,
  J.\ Math.\ Phys.\  {\bf 11}, no. 3, 941 (1970).
  
  \bibitem{Balakin} 
  A.~B.~Balakin and J.~P.~S.~Lemos,
  Class.\ Quant.\ Grav.\  {\bf 22}, 1867 (2005).
  
  \bibitem{Hehl} 
  F.~W.~Hehl and Y.~N.~Obukhov,
  Lect.\ Notes Phys.\  {\bf 562}, 479 (2001)
  
  
  \bibitem{Turner} 
  E.~W.~Kolb and M.~S.~Turner,
  Front.\ Phys.\  {\bf 69}, 1 (1990).
  
  \bibitem{Turner2} 
  M.~S.~Turner and L.~M.~Widrow,
  Phys.\ Rev.\ D {\bf 37}, 2743 (1988).
  
  \bibitem{Mazzitelli} 
  E.~A.~Calzetta, A.~Kandus and F.~D.~Mazzitelli,
  Phys.\ Rev.\ D {\bf 57}, 7139 (1998).
  
  \bibitem{Mazzitelli2} 
  F.~D.~Mazzitelli and F.~M.~Spedalieri,
  Phys.\ Rev.\ D {\bf 52}, 6694 (1995)
  
  \bibitem{Capozziello:1999uwa} 
  S.~Capozziello and G.~Lambiase,
  Gen.\ Rel.\ Grav.\  {\bf 31}, 1005 (1999).
  
  \bibitem{Campanelli:2008qp} 
  L.~Campanelli, P.~Cea, G.~L.~Fogli and L.~Tedesco,
  Phys.\ Rev.\ D {\bf 77}, 123002 (2008).
  
  \bibitem{novello} 
  M.~Novello, L.~A.~R.~Oliveira and J.~M.~Salim,
  Class.\ Quant.\ Grav.\  {\bf 13}, 1089 (1996).
  
  \bibitem{Iorio:2013ifa} 
  A.~Iorio and G.~Lambiase,
  Phys.\ Rev.\ D {\bf 90}, no. 2, 025006 (2014).
  
  \bibitem{Lambiase:2005kb} 
  G.~Lambiase,
  Phys.\ Rev.\ D {\bf 72}, 087702 (2005).
  
  \bibitem{ovgun1} 
  A.~\"{O}vg\"{u}n,
  Eur.\ Phys.\ J.\ C {\bf 77}, no. 2, 105 (2017)
  
  
  \bibitem{ovgun2} 
  A.~\"{O}vg\"{u}n, G.~Leon, J.~Magana and K.~Jusufi,
 Eur.\ Phys.\ J.\ C {\bf 78}, no. 6, 462 (2018).
  
  
  \bibitem{ovgun3} 
  G.~Otalora, A.~\"{O}vg\"{u}n, J.~Saavedra and N.~Videla,
JCAP {\bf 1806}, no. 06, 003 (2018).
  
  
  \bibitem{Weyl1}
  A.~Ritz and J.~Ward,
  Phys.\ Rev.\ D {\bf 79}, 066003 (2009).
  
\bibitem{Drummond} 
  I.~T.~Drummond and S.~J.~Hathrell,
  Phys.\ Rev.\ D {\bf 22}, 343 (1980).
  
  \bibitem{Dereli1} 
  T.~Dereli and O.~Sert,
  Eur.\ Phys.\ J.\ C {\bf 71}, 1589 (2011).
  
  \bibitem{Solanki} 
  S.~K.~Solanki {\it et al.},
  Phys.\ Rev.\ D {\bf 69}, 062001 (2004).
  

  
  \bibitem{Wu:2010vr} 
  J.~P.~Wu, Y.~Cao, X.~M.~Kuang and W.~J.~Li,
  Phys.\ Lett.\ B {\bf 697}, 153 (2011).
  
  
  \bibitem{Momeni:2011ca} 
  D.~Momeni and M.~R.~Setare,
  Mod.\ Phys.\ Lett.\ A {\bf 26}, 2889 (2011).
  
  
  \bibitem{Momeni:2012uc} 
  D.~Momeni, M.~R.~Setare and R.~Myrzakulov,
  Int.\ J.\ Mod.\ Phys.\ A {\bf 27}, 1250128 (2012).
  
  \bibitem{Momeni:2012ab} 
  D.~Momeni, N.~Majd and R.~Myrzakulov,
  EPL {\bf 97}, no. 6, 61001 (2012).
  
  \bibitem{Ma:2011zze} 
  D.~Z.~Ma, Y.~Cao and J.~P.~Wu,
  Phys.\ Lett.\ B {\bf 704}, 604 (2011).
  
  \bibitem{Roychowdhury:2012hp} 
  D.~Roychowdhury,
  Phys.\ Rev.\ D {\bf 86}, 106009 (2012).
  
  \bibitem{Zhao:2012kp} 
  Z.~Zhao, Q.~Pan and J.~Jing,
  Phys.\ Lett.\ B {\bf 719}, 440 (2013).

    
  
  
  \bibitem{Chen:2013ysa} 
  S.~Chen and J.~Jing,
  Phys.\ Rev.\ D {\bf 88}, 064058 (2013).
  
  \bibitem{Chen:2014zua} 
  S.~Chen and J.~Jing,
  Phys.\ Rev.\ D {\bf 90}, no. 12, 124059 (2014)
  
  
  \bibitem{Chen:2015bva} 
  J.~Jing, S.~Chen and Q.~Pan,
  Annals Phys.\  {\bf 367}, 219 (2016)
  

\bibitem{Chen:2015cpa} 
  S.~Chen and J.~Jing,
  JCAP {\bf 1510}, no. 10, 002 (2015).
  
\bibitem{Lu:2016gsf} 
  X.~Lu, F.~W.~Yang and Y.~Xie,
  Eur.\ Phys.\ J.\ C {\bf 76}, no. 7, 357 (2016).
  
\bibitem{Huang:2016qnl} 
  Y.~Huang, S.~Chen and J.~Jing,
  Eur.\ Phys.\ J.\ C {\bf 76}, no. 11, 594 (2016).
  
\bibitem{Chen:2016hil} 
  S.~Chen, S.~Wang, Y.~Huang, J.~Jing and S.~Wang,
  Phys.\ Rev.\ D {\bf 95}, no. 10, 104017 (2017).
  
\bibitem{Li:2017zsb} 
  G.~Li and X.~M.~Deng,
  Annals Phys.\  {\bf 382}, 136 (2017).
  
\bibitem{Cao:2018lrd} 
  W.~G.~Cao and Y.~Xie,
  Eur.\ Phys.\ J.\ C {\bf 78}, no. 3, 191 (2018).
  
  

  
\bibitem{Bluhm:2004ep} R.~Bluhm and V.~A.~Kostelecky, 
Phys.\ Rev.\ D \textbf{71}, 065008 (2005).



\bibitem{Bertolami:2005bh} O.~Bertolami and J.~Paramos, 
Phys.\ Rev.\ D \textbf{72}, 044001 (2005).



\bibitem{Bluhm:2008yt} R.~Bluhm, N.~L.~Gagne, R.~Potting and A.~Vrublevskis, 
Phys.\ Rev.\ D \textbf{77}, 125007 (2008) Erratum: [Phys.\ Rev.\ D \textbf{79%
}, 029902 (2009)].



\bibitem{Seifert:2009gi} M.~D.~Seifert, 
Phys.\ Rev.\ D \textbf{81}, 065010 (2010).






\bibitem{Escobar:2017fdi} C.~A.~Escobar and A.~Martin-Ruiz, 
Phys.\ Rev.\ D \textbf{95}, no. 9, 095006 (2017).

\bibitem{namb} M. H Dickinson,F. O Lehmann, S. P. Sane, Science \textbf{284}
(5422): 1954 (1999).

\bibitem{Kostelecky:1989jw} V.~A.~Kostelecky and S.~Samuel, 
Phys.\ Rev.\ D \textbf{40}, 1886 (1989).

\bibitem{Kostelecky:2010ze} A.~V.~Kostelecky and J.~D.~Tasson, 
Phys.\ Rev.\ D \textbf{83}, 016013 (2011).




  
    \bibitem{bumblebee} 
  R.~Casana, A.~Cavalcante, F.~P.~Poulis and E.~B.~Santos,
  Phys.\ Rev.\ D {\bf 97}, no. 10, 104001 (2018).

  
  \bibitem{Kostelecky:1989jp} 
  V.~A.~Kostelecky and S.~Samuel,
  Phys.\ Rev.\ Lett.\  {\bf 63}, 224 (1989).
  
  
  
  \bibitem{Kostelecky:1988zi} 
  V.~A.~Kostelecky and S.~Samuel,
  Phys.\ Rev.\ D {\bf 39}, 683 (1989).

\bibitem{LSB} O. Bertolami and C. Carvalho, J. Phys.: Conf. Ser.\textbf{\ 67}%
, 012009 (2007).

\bibitem{Maluf:2015hda} R.~V.~Maluf, J.~E.~G.~Silva and C.~A.~S.~Almeida, 
Phys.\ Lett.\ B \textbf{749}, 304 (2015).


\bibitem{Capelo:2015ipa} D.~Capelo and J.~Paramos, 
Phys.\ Rev.\ D \textbf{91}, no. 10, 104007 (2015).

\bibitem{Nascimento:2014vva} A.~F.~Santos, A.~Y.~Petrov, W.~D.~R.~Jesus and
J.~R.~Nascimento, 
Mod.\ Phys.\ Lett.\ A \textbf{30}, no. 02, 1550011 (2015).


\bibitem{Maluf:2014dpa} R.~V.~Maluf, C.~A.~S.~Almeida, R.~Casana and
M.~M.~Ferreira, Jr., 
Phys.\ Rev.\ D \textbf{90}, no. 2, 025007 (2014).

\bibitem{Paramos:2014mda} J.~Paramos and G.~Guiomar, 
Phys.\ Rev.\ D \textbf{90}, no. 8, 082002 (2014)



 
  \bibitem{Ovgun:2018xys} 
  A.~\"{O}vg\"{u}n, K.~Jusufi and I.~Sakalli,
   \url{arXiv:1804.09911 [gr-qc]}.
  
  
  
  
 \bibitem{gauss} C.F. Gauss, , Werke , 8 , K. Gesellschaft Wissenschaft. Göttingen (1900)
 
\bibitem{bonnet} O. Bonnet, J. Ecole Polytechnique , 19 (1848) pp. 1–146.
  
  \bibitem{GBT} Michiel Hazewinkel, ed. (2001), "Gauss-Bonnet theorem", Encyclopedia of Mathematics, Springer Science Business Media B.V., Kluwer Academic Publishers.
  
  \bibitem{gibbons2} 
  G.~W.~Gibbons,
  Phys.\ Lett.\ B {\bf 308}, 237 (1993).
  
  
\bibitem{gibbons1}
  G.~W.~Gibbons and M.~C.~Werner,
  Class.\ Quant.\ Grav.\  {\bf 25}, 235009 (2008).

\bibitem{werner} M.~C.~Werner,
  Gen.\ Rel.\ Grav.\  {\bf 44}, 3047 (2012).
  
  \bibitem{asahi1} A.~Ishihara, Y.~Suzuki, T.~Ono and H.~Asada,
  Phys.\ Rev.\ D {\bf 95}, no. 4, 044017 (2017).

\bibitem{Ono:2017pie} 
  T.~Ono, A.~Ishihara and H.~Asada,
  Phys.\ Rev.\ D {\bf 96}, no. 10, 104037 (2017).
  
  
\bibitem{Jusufiwh} 
  K.~Jusufi,
  Int.\ J.\ Geom.\ Meth.\ Mod.\ Phys.\  {\bf 14}, no. 12, 1750179 (2017).

\bibitem{Jusufi:2017mav} 
  K.~Jusufi and A.~\"{O}vg\"{u}n,
  Phys.\ Rev.\ D {\bf 97}, no. 2, 024042 (2018)
  


\bibitem{kimet2}  K.~Jusufi,
  Eur.\ Phys.\ J.\ C {\bf 76}, no. 6, 332 (2016).

\bibitem{kimet22}
  K.~Jusufi,
  Astrophys.\ Space Sci.\  {\bf 361}, no. 1, 24 (2016).

\bibitem{wh10}   K.~Jusufi,
  Int.\ J.\ Geom.\ Meth.\ Mod.\ Phys.\  {\bf 14}, no. 10, 1750137 (2017).

\bibitem{kimet1} K. Jusufi, M. C. Werner, A. Banerjee and A.  \"{O}vg\"{u}n ,
 Phys.\ Rev.\ D \textbf{95}, no. 10, 104012 (2017).

\bibitem{kaa} K. Jusufi, A. \"{O}vg\"{u}n and A. Banerjee, 
 Phys.\ Rev.\ D \textbf{96}, no. 8, 084036 (2017).

\bibitem{kimet3} K. Jusufi, I. Sakalli, A. \"{O}vg\"{u}n, 
 Phys.\ Rev.\ D \textbf{96}, no. 2, 024040 (2017).

\bibitem{aovgun} I. Sakalli and A. \"{O}vg\"{u}n, 
 Europhys.\ Lett.\ \textbf{118}, no. 6, 60006 (2017).
 
 \bibitem{aovgun1} 
  A.~\"{O}vg\"{u}n,
 Phys.\ Rev.\ D {\bf 98}, 044033 (2018).
  
  \bibitem{asada}   K.~Nakajima and H.~Asada,
  Phys.\ Rev.\ D {\bf 85}, 107501 (2012).
  
  
\bibitem{Crisnejo:2018uyn} 
  G.~Crisnejo and E.~Gallo,
  Phys.\ Rev.\ D {\bf 97}, no. 12, 124016 (2018).



\bibitem{Jusufi:2018jof} 
  K.~Jusufi, A.~\"{O}vg\"{u}n, J.~Saavedra, P.~A.~Gonzalez and Y.~Vasquez,
 Phys.\ Rev.\ D {\bf 97}, no. 12, 124024 (2018).




\bibitem{Jusufi:2018kmk} 
  K.~Jusufi, A.~\"{O}vg\"{u}n, A.~Banerjee and ˙ I.~Sakalli,
   \url{arXiv:1802.07680 [gr-qc]}.


\bibitem{Jusufi:2017drg} 
  K.~Jusufi, N.~Sarkar, F.~Rahaman, A.~Banerjee and S.~Hansraj,
  Eur.\ Phys.\ J.\ C {\bf 78}, no. 4, 349 (2018).
  
\bibitem{Jusufi:2017uhh} 
  K.~Jusufi and A.~\"{O}vg\"{u}n,
  Phys.\ Rev.\ D {\bf 97}, no. 6, 064030 (2018).
  

\bibitem{Jusufi:2017xnr} 
  K.~Jusufi, F.~Rahaman and A.~Banerjee,
  Annals Phys.\  {\bf 389}, 219 (2018).
  




\bibitem{Jusufi:2017vew} 
  K.~Jusufi and A.~\"{O}vg\"{u}n,
   \url{arXiv:1707.02824 [gr-qc]}.



\bibitem{Ishihara:2016vdc} 
  A.~Ishihara, Y.~Suzuki, T.~Ono, T.~Kitamura and H.~Asada,
  Phys.\ Rev.\ D {\bf 94}, no. 8, 084015 (2016).
  

\bibitem{Jusufi:2015laa} 
  K.~Jusufi,
  Astrophys.\ Space Sci.\  {\bf 361}, no. 1, 24 (2016)


\bibitem{Gibbons:2015qja} 
  G.~W.~Gibbons,
  Class.\ Quant.\ Grav.\  {\bf 33}, no. 2, 025004 (2016).
  
  
  

  
  
  \bibitem{eventhorizon}  \url{https://eventhorizontelescope.org}.


\bibitem{weinberg} S. Weinberg, \textit{Gravitation and Cosmology}
(New York: John Wiley \& Sons, 1972).


\bibitem{blackholes-wormholes1}   N.~Tsukamoto, T.~Harada and K.~Yajima,
  Phys.\ Rev.\ D {\bf 86}, 104062 (2012).


\bibitem{bao} D. Bao, S. Chern and Z. Shen, \textit{An Introduction
to Riemann-Finsler Geometry}. Springer, New York (2000)

\bibitem{Breton}
  N.~Breton,
  Class.\ Quant.\ Grav.\  {\bf 19}, 601 (2002).
  
  \bibitem{Einstein:1935tc} 
  A.~Einstein and N.~Rosen,
  Phys.\ Rev.\  {\bf 48}, 73 (1935).
  
  \bibitem{Beringer} J. Beringer et al. (Particle Data Group), Phys. Rev. D
\textbf{86}, 010001 (2012).



\end{thebibliography}
\end{document}